\documentclass[namedreferences]{SolarPhysics}
\usepackage[optionalrh]{spr-sola-addons} 
\usepackage{graphicx}        
\usepackage{color}           
\usepackage{url}             

\usepackage{solaheader}	
\newcommand{\solareceiveddate}{21 January 2009} 
\newcommand{\solaaccepteddate}{3 February 2009}

\begin{document}

\begin{article}

\begin{opening}

\title{Monte-Carlo Simulation of Solar Active-Region Energy}

\author{M.S.~\surname{Wheatland$^{1}$}}
\runningauthor{M. Wheatland}
\runningtitle{Monte-Carlo Simulation of Active Region Energy}

\institute{$^{1}$ School of Physics, University of Sydney, NSW 2006, 
Australia \\
email: \url{m.wheatland@physics.usyd.edu.au}}

\begin{abstract}
A Monte-Carlo approach to solving a stochastic jump transition
model for active-region energy 
(Wheatland and Glukhov, {\it Astrophys.\ J.} {\bf 494},
1998; Wheatland, {\it Astrophys.\ J.} {\bf 679}, 2008)
is described. The new method numerically solves the stochastic 
differential equation describing the model, rather than the 
equivalent master equation. This has the advantages of allowing 
more efficient numerical solution, the modelling of time-dependent 
situations, and investigation of details of event statistics.
The Monte-Carlo approach is illustrated by application to a 
Gaussian test case, and to the class of flare-like models presented 
in \inlinecite{2008ApJ...679.1621W}, which are steady-state models
with constant rates of energy supply, and power-law distributed jump 
transition rates. These models
have two free parameters: an index ($\delta $), which defines the 
dependence of the jump transition rates on active-region energy, and
a non-dimensional ratio ($\overline{r})$ of total flaring rate
to rate of energy supply. For $\overline{r}\ll 1$ the non-dimensional
mean energy $\langle \overline{E}\rangle$ of the active-region 
satisfies $\langle \overline{E}\rangle \gg 1$, resulting in a power-law
distribution of flare events over many decades in energy.
The Monte-Carlo method is used to explore the behavior of the 
waiting-time distributions for the flare-like models. The models 
with $\delta\neq 0$ are found to have waiting times which depart 
significantly from simple Poisson behavior when 
$\langle \overline{E}\rangle \gg 1$. The original model from 
\inlinecite{1998ApJ...494..858W}, with $\delta=0$ (no dependence of 
transition rates on active-region energy), is identified as being 
most consistent with observed flare statistics.
\end{abstract}
\end{opening}

\section{Introduction}
     \label{sec:Introduction} 

Solar flares are explosive events in the solar corona, involving the
release of energy stored in active-region magnetic
fields. Active regions are dynamic, evolving in time due to the
emergence and submergence of magnetic flux from the sub-photosphere,
stressing by photospheric motions, and the occurrence of flares. It
is of interest to understand the dynamical energy balance of active
regions.

Flares exhibit a wide range of energies. The largest flares may involve
the release of up to $10^{27}\,{\rm J}$ of energy, and are associated
with large-scale expulsions of material from the corona (Coronal Mass
Ejections, or CMEs), involving comparable energies. Small flare-like
events are observed down to the limits of observation and are difficult
to distinguish from a variety of small-scale solar activity. The
distribution of flare energies follows a power-law distribution
\cite{1991SoPh..133..357H, 1993AdSpR..13..179C,1998ApJ...497..972A}.
Specifically, the frequency-energy distribution ${\cal N}(E)$, 
{\it i.e.} the number of flares observed per unit time and per unit 
energy ($E$), obeys
\begin{equation}\label{eq:obs_pl} 
{\cal N}(E)=AE^{-\gamma},
\end{equation} 
where the factor $A$ is a (time-dependent) measure of the
total flaring rate, and $\gamma\approx 1.5$. This distribution is often
constructed for flares from all active regions present on the Sun over
some period of time, but it also appears to apply to individual active
regions \cite{2000ApJ...532.1209W}, which suggests that the power law is
fundamental to the flare mechanism. A popular model explaining the power
law is the avalanche model
\cite{1991ApJ...380L..89L,2001SoPh..203..321C}, in which the magnetic
field in the corona is assumed to be in a self-organized critical state,
and subject to avalanches of small-scale reconnection events. The 
distribution~(\ref{eq:obs_pl}) must have an upper roll-over,
to ensure that the total energy released in flares is finite
\cite{1997ApJ...475..338K,2007ApJ...663L..45H}.

The mechanisms causing flares are not well understood, and flares appear
to occur randomly in time, although certain properties of active regions
correlate with flaring ({\it e.g.} \opencite{2000ApJ...540..583S}; 
\opencite{1990SoPh..125..251M}; \opencite{2007ApJ...661L.109G};
\opencite{2007ApJ...655L.117S}). 
Correspondingly, the prediction of flares is in its infancy: the methods 
used are probabilistic and are rather inaccurate at predicting the
occurrence of large flares, which are rare but which strongly influence
our local space weather \cite{2005SpWea...307003W,2007SpWea...509002B,
2008ApJ...688L.107B}. 

The occurrence of flares in time may be investigated via a
second observable distribution, the flare waiting-time distribution, or
the distribution of times between events (this distribution is also
referred to as the ``interval distribution''). Determinations of flare
waiting-time distributions have given varied results
\cite{1993Ap&SS.208...99P,1994PhDT........51B,1998ApJ...509..448W,
1999PhRvL..83.4662B,2001ApJ...555L.133L,
2001JGR...10629951M,2001SoPh..203...87W,
2002ApJ...574..434M,2008SoPh..248...85K}, suggesting that the observed
distribution may depend on the particular active region, on time, and
that it also may be influenced by event definition and selection 
procedures \cite{2001SoPh..203...87W,2005A&A...436..355B,
2005PhRvL..95r1102P,2006PhRvL..96e1103B}. For some active regions, the
distribution is consistent with a simple Poisson process, {\it i.e.}
independent events occurring at a constant mean rate 
\cite{2001JGR...10629951M}, and
the corresponding waiting-time distribution is exponential. 
Other active regions exhibit time-variation in the flaring process, and 
flare occurrence may be approximated by a piecewise-constant, or more 
generally time-varying Poisson process, in which case the distribution 
is a sum or integral over exponentials \cite{2001SoPh..203...87W}. On 
longer time scales, the distribution exhibits a power-law tail for events 
from the whole Sun \cite{1999PhRvL..83.4662B}.
Some authors have also argued that the process is fundamentally 
non-Poissonian ({\it e.g.} \opencite{2001ApJ...555L.133L}), although the
arguments neglect the role of time-dependence.

The energy balance of active regions presents a puzzle, because of the 
large drops in energy due to large solar flares. In the following
we consider a simple ``black box'' approach to modelling the free energy 
of an active region (the energy available to power flares), an approach 
that goes back to \inlinecite{1978ApJ...222.1104R}.
Energy is assumed to be continuously supplied to the active region 
by some external mechanism. The energy is stored locally in the corona, 
and some is released at particular times in flares. The flaring 
process is considered to be stochastic, whereas energy supply is 
deterministic. Flares are treated as point processes in time, {\it i.e.} 
they occur at one instant in time, and they involve 
jump transitions (discontinuous changes) in energy. A general 
model of this kind was presented in \inlinecite{1998ApJ...494..858W} 
and further developed in \inlinecite{2008ApJ...679.1621W}. 
The general model is based on a master equation for the probability 
distribution [$P(E,t)$] for the free energy ($E$) of an active region 
at time $t$. The free parameters in the model are a prescribed energy 
supply rate to the system [$\beta (E,t)$], and prescribed transition 
rates [$\alpha (E,E^{\prime},t)$], describing the rate of jumps from 
energy $E$ to energy $E^{\prime}\leq E$. These two rate functions
may have any functional form. In \inlinecite{1998ApJ...494..858W} it 
was argued that the energy-supply rate should not depend on the energy 
($E$) of the system, since active regions are driven externally, and 
hence a constant energy supply rate is appropriate. It was also argued 
that, to produce an appropriate power-law flare frequency-energy 
distribution, transition rates of the form 
$\alpha (E,E^{\prime})\sim (E-E^{\prime})^{-\gamma}$
are required (in the steady state). \inlinecite{1998ApJ...494..858W} 
and \inlinecite{2008ApJ...679.1621W} investigated the model by solving 
the master equation in the steady state, for these ``flare-like''
choices for the free parameters. In \inlinecite{1998ApJ...494..858W}
the emphasis was on the basic model and the arguments for the 
appropriate flare-like choices. A model was constructed that could
reproduce power-law behavior in the flare frequency-energy 
distribution over an arbitrary number of decades in energy, up
to a high energy roll-over set by the decline of $P(E)$ at large
energy. Hence the flare-like model was confirmed to reproduce this
aspect of flare statistics. In \inlinecite{2008ApJ...679.1621W}
the waiting-time distributions for the model were also considered,
using theory for jump transition processes presented for the first
time by \inlinecite{2007PhRvE..75a1119D}. It was found that the 
\inlinecite{1998ApJ...494..858W} model produces an essentially 
Poisson (exponential) waiting-time distribution. A modified
model was also considered, involving an 
$\alpha (E,E^{\prime})\sim E (E-E^{\prime})^{-\gamma}$ form
for the transition rates. This model also reproduced the power-law
frequency-energy distribution, but exhibited some departure from 
a simple Poisson waiting-time distribution.

Master equations may be represented by an equivalent stochastic 
differential equation \cite{1992sppc.book.....V,Gardiner2004}, which
provides a complementary approach to the problem at hand. Stochastic 
DEs are amenable to solution by Monte-Carlo methods, which in general 
are simple to numerically implement. 
This paper describes a Monte-Carlo approach to solving the stochastic 
model for solar active-region energy presented in 
\inlinecite{1998ApJ...494..858W} and
\inlinecite{2008ApJ...679.1621W}. The Monte-Carlo approach has 
specific advantages over the master-equation approach: it is 
computationally more efficient, it permits more general modelling, 
in particular the modelling solution of time-dependent problems, and
it generates an ensemble of flare events and hence permits detailed
investigation of event statistics. 

The layout of the paper is as follows. The master-equation approach of
\inlinecite{1998ApJ...494..858W} and \inlinecite{2008ApJ...679.1621W}
is briefly reiterated in Section~\ref{sec:Master-General}, and
the flare-like choices for the model are explained in 
Section~\ref{sec:Master-Flare-like}. 
The stochastic DE approach to the problem is then presented in 
Section~\ref{sec:SDE-General}, and illustrated by application to a
Gaussian test case  (Section~\ref{sec:SDE-Gauss}), and to the 
flare-like cases from \inlinecite{2008ApJ...679.1621W} 
(Section~\ref{sec:SDE-Flare-like}), including comparison of a 
Monte-Carlo solution with direct numerical solution of the master 
equation. Section~\ref{sec:SDE-Flare-like-WTD} presents a 
Monte-Carlo-based investigation of the variation of the waiting-time
distribution for the flare-like models, 
and Section~\ref{sec:Conclusions} presents conclusions.

\section{Master Equation Approach} 
\label{sec:Master}      

\subsection{GENERAL APPROACH} 
\label{sec:Master-General}

To begin we briefly reiterate the master-equation formulation of the 
model, following \inlinecite{1998ApJ...494..858W} and 
\inlinecite{2008ApJ...679.1621W}. The energy [$E=E(t)$] of an active 
region is assumed to be a stochastic variable which evolves in time due 
to deterministic energy input at a rate $\beta (E,t)$, as well as due to 
jumps downwards in energy (flares) at random times and of random sizes, 
described by transition rates $\alpha (E,E^{\prime},t)$. These are the 
rate for jumps per unit energy from $E$ to $E^{\prime}$ at time $t$. 
The probability distribution [$P(E,t)$] for the energy of the system is 
given by the solution to the master equation 
\begin{eqnarray}\label{eq:master}
\frac{\partial P(E,t)}{\partial t} &=&
-\frac{\partial }{\partial E}\left[\beta (E,t) P(E,t)\right]
-\lambda (E,t)P(E,t)
\nonumber \\ &+&\int_E^{\infty}P(E^{\prime},t)\alpha
(E^{\prime},E,t){\mathrm d}E^{\prime},
\end{eqnarray}
where
\begin{equation}\label{eq:lambda}
\lambda (E,t)=\int_0^{E}\alpha
(E,E^{\prime},t){\mathrm d}E^{\prime}
\end{equation}
is the total rate of flaring at time $t$, assuming the system has energy 
$E$. [A time dependence has been included in the transition rate, by 
contrast with \inlinecite{2008ApJ...679.1621W}.] Two other quantities
of interest are the mean total rate of transitions (in the average over
energy)
\begin{equation}\label{eq:lambda_bar}
\langle\lambda \rangle =\int_0^{\infty}\lambda(E,t)P(E,t){\mathrm d}E
\end{equation}
and the mean energy of the system
\begin{equation}\label{eq:ebar}
\langle E \rangle =\int_0^{\infty}E P(E,t){\mathrm d}E.
\end{equation}

As noted in Section~\ref{sec:Introduction}, two observable flare 
distributions are the flare frequency-energy distribution and the 
waiting-time distribution. The model frequency-energy distribution 
is given by
\begin{equation}\label{eq:ffe}
{\cal N}(E,t)=\int_{E}^{\infty}P(E^{\prime},t)\alpha
(E^{\prime},E^{\prime}-E,t){\mathrm d}E^{\prime}.
\end{equation}
\inlinecite{2007PhRvE..75a1119D} showed how to obtain the
distribution of waiting times ($\tau$) for jump transition models 
in the steady state ($\partial /\partial t = 0$). In 
\inlinecite{2008ApJ...679.1621W} that theory was applied to
Equation~(\ref{eq:master}) to yield the model waiting-time distribution 
$p_{\tau}(\tau )$. The details of the derivation are given in
\inlinecite{2008ApJ...679.1621W}.

In \inlinecite{1998ApJ...494..858W} and \inlinecite{2008ApJ...679.1621W}
the master equation was numerically solved in the steady state, for 
flare-like choices of $\beta (E)$ and $\alpha (E,E^{\prime})$ (the 
choices are explained in Section~\ref{sec:Master-Flare-like}). The
methods of solution involved discretising the energy as a set of values 
$E_i$ ($i=1,2,3,...,N$), in which case the master equation represents
a system of $N$ linear equations in $N$ unknowns $P_i=P(E_i)$. Solution 
of the linear system was performed either by relaxation 
\cite{1998ApJ...494..858W}, or by back substitution 
\cite{2008ApJ...679.1621W}. One disadvantage of these methods
is that the energy may span many decades in energy, in which case 
a large value of $N$ is required.

\subsection{FLARE-LIKE CHOICES} 
\label{sec:Master-Flare-like}

In \inlinecite{1998ApJ...494..858W}, the master equation was solved in 
the steady state ($\partial /\partial t = 0$) for the choices 
$\beta (E) = \beta_0$, a constant, and 
\begin{equation}\label{eq:pl_alpha_wg98}
\alpha (E,E^{\prime})=\alpha_0 (E-E^{\prime})^{-\gamma} \theta
(E-E^{\prime}-E_c),
\end{equation}
where $\alpha_0$ is a constant, 
$E_c$ is a low-energy cutoff, and 
$\theta (x)$ is the step function. In \inlinecite{2008ApJ...679.1621W}
Equation~(\ref{eq:pl_alpha_wg98}) was generalised to include an 
additional dependence on the initial energy $E$:
\begin{equation}\label{eq:pl_alpha_whe08}
\alpha (E,E^{\prime})=\alpha_0E^{\delta} (E-E^{\prime})^{-\gamma} 
\theta (E-E^{\prime}-E_c),
\end{equation}
where $\delta$ is a constant.
The problem was solved for Equation~(\ref{eq:pl_alpha_whe08}) with 
$\delta = 0$ and $\delta = 1$, and for $\beta (E) =\beta_0$.

The physical motivations for these choices is briefly mentioned
in Section~\ref{sec:Introduction}, but is worth discussing in
more detail. 
Concerning the energy-supply rate, the physical aspect is that the 
rate does not depend on the energy of the system. This is appropriate 
for a system that is externally driven, and for which there is no 
back reaction of the system on the driver. The picture for the Sun 
is that energy supply comes from below ({\it i.e.} from the 
sub-photosphere), {\it via} photospheric flows which cause new fields 
to emerge into the corona, and twist existing coronal fields. The rate 
at which this occurs is determined by flow patterns in the 
sub-photosphere. The sub-photosphere is very dense, so it is unlikely 
that the corona can influence the rate of supply of energy, assuming 
this picture of energy supply is correct. Of course, the energy supply 
rate may depend on time, and the choice of a constant supply rate is 
a simplification. We will return to the question of time-dependent 
driving in Section~\ref{sec:Conclusions}.

Concerning the functional forms for the transition rates, first 
note that substituting Equation~(\ref{eq:pl_alpha_whe08}) into
Equation~(\ref{eq:ffe}) leads to the flare frequency-energy 
distribution
\begin{equation}\label{eq:ffe_pl_alpha}
{\cal N}(E)=\alpha_0 E^{-\gamma}
\int_{E}^{\infty}\left(E^{\prime}\right)^{\delta}
P(E^{\prime}){\mathrm d}E^{\prime},
\end{equation}
for $E\geq E_c$. It follows that the frequency-energy distribution 
is a power law with index $\gamma$ up to energies $E$ at which 
$P(E)$ becomes very small. This is consistent with the observed 
power-law frequency-energy distribution~(\ref{eq:obs_pl}), and the 
physical requirement that 
the frequency-energy distribution rolls over at large energies 
(to ensure the total mean rate of energy release in flares is 
finite). An estimate of the energy for departure from power-law 
behavior is provided by the mean energy, which may be approximated 
by~\cite{1998ApJ...494..858W}
\begin{equation}\label{eq:ebar_pl_alpha}
\langle E\rangle \approx
\left(\frac{2-\gamma}{\alpha_0/\beta_0}
\right)^{1/(\delta+2-\gamma )}.
\end{equation}
In principle, other functional choices leading to power-law behavior 
are possible, although it has proven difficult to identify alternative
solutions with an energy-supply rate independent of energy
that produce a power-law form for ${\cal N} (E)$. 

Another motivation for the choices~(\ref{eq:pl_alpha_wg98}) 
and~(\ref{eq:pl_alpha_whe08}) for the transition rates comes from 
consideration of avalanche type 
models~\cite{1991ApJ...380L..89L,2001SoPh..203..321C}. In these
models the volume involved in flaring is the set of unstable sites
which trigger one another during the flare ``avalanche.'' The volume 
of this region is found to be scale free, {\it i.e.} power-law 
distributed, and is fractal in shape. Assuming the volume of the
region is proportional to the energy released, this implies a form
$\sim (E-E^{\prime})^{-\gamma}$ for the probability per unit time
of a transition from energy $E$ to $E^{\prime}$, assuming flares 
occur at a constant rate per unit time. Hence the flare-like choices
for the master equation may correspond to the avalanche model, although
the detailed relationship between the two pictures remains to be
worked out.

Some support for these choices is provided by the resulting
waiting-time distributions. The numerical solutions for $P(E)$ in 
\inlinecite{2008ApJ...679.1621W} lead to
waiting-time distributions which are approximately exponential. 
This may be
understood by noting that substituting 
Equation~(\ref{eq:pl_alpha_whe08}) into Equation~(\ref{eq:lambda}) 
gives the total rate for flaring
\begin{equation}\label{eq:lambda_W08}
\lambda (E,t)=\left\{
  \begin{array}{ll}
    \alpha_0 E^{\delta} 
    \left(E_c^{-\gamma+1}-E^{-\gamma+1}\right)/(\gamma-1)
      & \mbox{if $E\geq E_c$,} \\
      0 & \mbox{else.}
  \end{array}\right.
\end{equation}
For $E\gg E_c$ we have
\begin{equation}\label{eq:lambda_W08_largeE}
\lambda (E) \approx  
  \frac{\alpha_0}{\gamma-1} E_c^{-\gamma+1} E^{\delta}.
\end{equation}
For $\delta = 0$, Equation~(\ref{eq:lambda_W08_largeE}) implies 
$\lambda (E)$ is constant (independent of $E$). For $\delta =1$ we have 
$\lambda (E) \propto E$, and hence the mean rate may be approximately 
constant, provided $P(E)$ is non-zero only over a fairly limited range 
in $E$. Hence the total rate of flaring may be approximately constant 
for $\delta =0$ and $\delta =1$, consistent with a simple Poisson 
process. As discussed in 
Section~\ref{sec:Introduction}, this is compatible with observed flare 
statistics in active regions for which the rate does not vary in time 
\cite{2001JGR...10629951M,2001SoPh..203...87W}.
 
\section{Stochastic DE Approach} 
\label{sec:SDE}      

\subsection{GENERAL APPROACH} 
\label{sec:SDE-General}    

Following \inlinecite{2007PhRvE..75a1119D}, the master 
Equation~(\ref{eq:master}) is equivalent to the stochastic differential 
equation
\begin{equation}\label{eq:stoch_de}
\frac{{\mathrm d}E}{{\mathrm d}t}=\beta (E,t)-\Lambda (E,t)
\end{equation}
where 
\begin{equation}\label{eq:Lambda}
\Lambda (E,t)=\sum_{i=1}^{N(t)}\Delta E_i\delta (t-t_i)
\end{equation}
describes the loss in energy due to flaring, with 
$\delta (x)$ being the delta function, $N(t)$ being the number of 
events up to time $t$, and with the event times $t_i$ defined
by a ``state-dependent'' Poisson process with occurrence rate 
$\lambda (E,t)$. The jump amplitudes $\Delta E$ follow the distribution 
$h (\Delta E,E,t)$, defined by
\begin{equation}\label{eq:h}
\alpha (E,E-\Delta E,t )=\lambda (E,t)h (\Delta E,E,t),
\end{equation}
so that
\begin{equation}\label{eq:hnorm}
\int_0^E h(\Delta E,E,t){\mathrm d}(\Delta E)=1.
\end{equation}

The ODE~(\ref{eq:stoch_de}) may be solved
in the following way. First, choose a start-energy $E_s$ at
time $t_s$. The energy of the system evolves deterministically
from this time up until the first jump at time $t_e=t_s+\tau$, 
where $\tau$ is a waiting time. The waiting time corresponds to a 
time-dependent Poisson process with a rate 
$\lambda \left[E(t),t\right]$. To evaluate this rate, note that the 
energy during the deterministic trajectory obeys
\begin{equation}\label{eq:stoch_de_nojumps}
\frac{{\mathrm d}E}{{\mathrm d}t}=\beta (E,t).
\end{equation}
Solving Equation~(\ref{eq:stoch_de_nojumps}) with the initial
condition $E=E_s$ at $t=t_s$ defines the time history $E^{\ast}(t)$ 
for the energy, and this 
together with $\lambda = \lambda (E,t)$
defines the rate $\lambda \left[E^{\ast}(t),t\right]$ prior to a 
jump. A waiting time may be generated for this rate by finding the 
root of the monotonic function \cite{2006SoPh..238...73W}:
\begin{equation}\label{eq:Ftau}
F(\tau)=\ln (1-u) +\int_{t_s}^{t_s+\tau}\lambda 
  \left[E^{\ast}(t),t \right]{\mathrm d}t,
\end{equation}
where $u$ is a uniform deviate (a uniformly-distributed number in
the range $0\leq u <1$). If numerical root finding is required, then
Newton-Raphson is a suitable method ({\it e.g.} Press {\it et al.} 
1992), and in that case it is worth noting that
\begin{equation}\label{eq:Ftauprime}
F^{\prime}(\tau )=\lambda\left[E^{\ast}(t_e ),t_e\right],
\end{equation}
where $t_e=t_s+\tau$.

Once the waiting time $\tau$ has been generated, the jump
may be simulated. The energy before the jump is 
$E_e=E^{\ast}(t_e )$,
the end-energy of the deterministic trajectory. The size $\Delta E$ of
the jump may be determined by generating a random variable from the
distribution $h(\Delta E,E_e,t_e)$, which is defined by 
Equation~(\ref{eq:h}). A value $\Delta E$ may be obtained by the usual 
technique of transforming a uniform deviate to a random variable from 
the required distribution \cite{1992nrca.book.....P}. 
Once $\Delta E$ 
is calculated, a new start energy $E_s=E_e-\Delta E$ is specified at 
time $t_e$ just after the jump, and the whole process may then 
be repeated. This procedure can be repeated an arbitrary number of 
times, to give a simulated time history of energy $E(t)$ for the system 
over an arbitrary number of jumps.
Provided Equations~(\ref{eq:stoch_de_nojumps}) and~(\ref{eq:Ftau}) 
are straightforward to evaluate, the process involves relatively 
little computational expense.

\subsection{GAUSSIAN TEST CASE} 
\label{sec:SDE-Gauss}   

To illustrate the method, consider the ``Gaussian'' test case discussed
in \inlinecite{1998ApJ...494..858W} and 
\inlinecite{2008ApJ...679.1621W}, namely the
case with $\beta (E,t)=\beta_0$ and 
$\alpha (E,E^{\prime},t)=\alpha_0$ 
(where $\alpha_0$ and $\beta_0$ are constants). In that case the 
solution to the steady state master equation is
\begin{equation}\label{eq:PE_gauss}
P(E)=aE{\mathrm e}^{-\frac{1}{2}aE^2},
\end{equation}
with $a=\alpha_0/\beta_0$,
and from Equation~(\ref{eq:ffe}) the frequency-energy distribution for
jumps is also a Gaussian:
\begin{equation}
{\cal N}(E)=\alpha_0{\mathrm e}^{-\frac{1}{2}aE^2}.
\end{equation}
From Equation~(\ref{eq:lambda}) the total rate of events is 
$\lambda (E)=\alpha_0 E$, and from Equation~(\ref{eq:lambda_bar}) the 
mean total rate is
$\langle \lambda \rangle =\left(\alpha_0\beta_0\right)^{1/2}$. 
Using the method outlined in \inlinecite{2008ApJ...679.1621W},
the waiting-time distribution is also a Gaussian:
\begin{equation}\label{eq:Ptau_gauss}
p_{\tau}(\tau)=\left(\frac{2\alpha_0\beta_0}{\pi}\right)^{1/2}
{\mathrm e}^{-\frac{1}{2}\alpha_0\beta_0\tau^2}.
\end{equation}

To simulate the Gaussian test case using the Monte-Carlo approach, 
note that the solution to Equation~(\ref{eq:stoch_de_nojumps}) with
$\beta (E,t)=\beta_0$ and with starting energy $E_s$ at time $t_s$ is
\begin{equation}\label{eq:Et_const_beta0}
E^{\ast}(t)=E_s+\beta_0 \left( t-t_s\right),
\end{equation}
and we have $\lambda (E) = \alpha_0 E$, so
\begin{equation}
\lambda[E^{\ast}(t)]=\alpha_0\left[E_s+\beta_0\left(t-t_s\right)\right].
\end{equation}
Equation~(\ref{eq:Ftau}) evaluates to
\begin{equation}
F(\tau)=\ln (1-u) + \alpha_0E_s\tau
  +\frac{1}{2}\alpha_0\beta_0\tau^2,
\end{equation}
and taking the positive root of this quadratic function gives
\begin{equation}
\tau=\frac{E_s}{\beta_0}
  \left[\left(1+\frac{2}{aE_s^2}\ln \frac{1}{1-u}\right)^{1/2}-1\right].
\end{equation}
For this case the distribution of jump amplitudes, from
Equation~(\ref{eq:h}), is given by
\begin{equation}
h (\Delta E,E) = \frac{1}{E}
\end{equation}
for $0\leq \Delta E\leq E$. The jump energies are uniformly distributed 
on $(0,E)$, and a jump may be generated from a uniform deviate $u$ 
using $\Delta E = Eu$. Finally, the mean energy
\begin{equation}\label{eq:Emean_gauss}
\langle E\rangle = \left(\frac{\pi}{2a}\right)^{1/2}
\end{equation}
provides a suitable starting energy for simulation.

\begin{figure}[here]
\centerline{\includegraphics[width=0.75\textwidth]{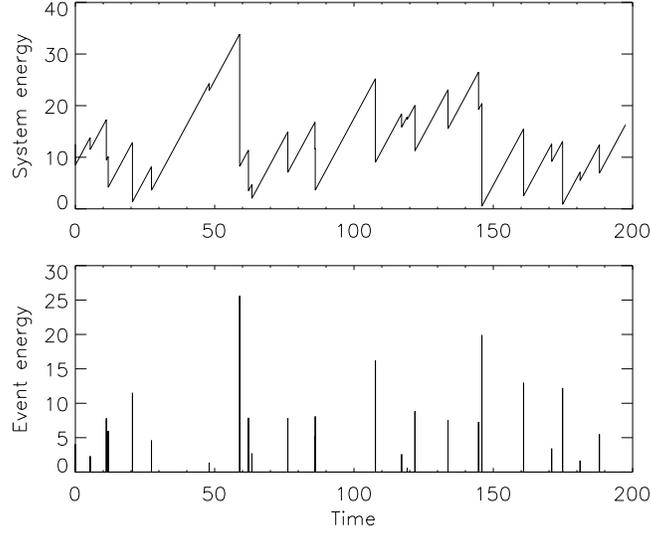}}
\caption{Monte-Carlo solution of the Gaussian test case with 
$\alpha_0=0.01$ and $\beta_0=1$, with 25 jump transitions and
waiting times. The upper panel shows the system energy {\it versus} 
time, and the lower panel shows the event energies {\it versus} time.}
\label{fig:fig1}
\end{figure}

Figure~\ref{fig:fig1} illustrates a Monte-Carlo solution of the 
Gaussian test case using 
Equations~(\ref{eq:Et_const_beta0})\,--\,(\ref{eq:Emean_gauss}),
for the choices $\alpha_0=0.01$ and $\beta_0=1$. The upper panel in the 
figure shows the time history of system energy for the simulation for 25 
jumps, and the lower panel shows the corresponding event energies 
{\it versus} time. The energy of the system grows linearly with time between 
jumps, and the jumps occur with a rate which increases linearly with 
energy. The jump sizes are uniformly distributed, up to the current
energy of the system.

\begin{figure}[here]
\centerline{\includegraphics[width=1\textwidth]{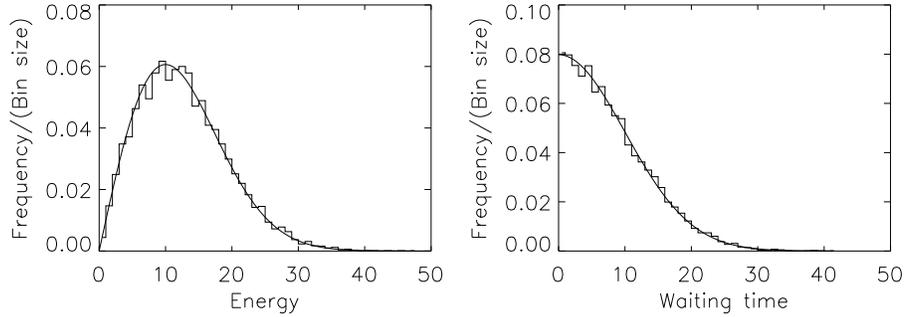}}
\caption{Monte-Carlo solution of the Gaussian test case, involving
 5000 jump transitions, and parameter values $\alpha_0=0.01$,
 $\beta_0=1$. The histogram in the left-hand panel shows the 
 distribution of system energy, and the histogram in the right-hand
 panel shows the distribution of waiting times. The corresponding
 analytic results are shown in the two panels by solid curves.}
\label{fig:fig2}
\end{figure}

Figure~\ref{fig:fig2} illustrates the Monte-Carlo solution of the
Gaussian test case with the same parameters ($\alpha_0=0.01$ and 
$\beta_0=1$) for 5000 waiting times and jumps, and compares the
results with the analytic expressions. The left-hand panel
shows the histogram of the system energy, and the right-hand panel 
shows the histogram of waiting times. The corresponding analytic
distributions $P(E)$ and $p_{\tau}(\tau)$, given by 
Equations~(\ref{eq:PE_gauss}) and~(\ref{eq:Ptau_gauss})
respectively, are shown by the solid curves. The histogram of system
energy was obtained by sampling the simulated time history of energy 
$E(t)$ at 5000 random times, uniformly distributed over the duration 
of the simulation. These results illustrate the application of the 
Monte-Carlo approach, and confirm that it reproduces the analytic 
results.

\subsection{FLARE-LIKE CASES} 
\label{sec:SDE-Flare-like}

Next we consider the flare-like cases corresponding to 
Equation~(\ref{eq:pl_alpha_whe08}), from 
\inlinecite{2008ApJ...679.1621W}. For those case the total rate 
of flaring is given by Equation~(\ref{eq:lambda_W08}),
and the energy supply rate is $\beta=\beta_0$, a constant.
If the last event is at $t=t_s$, when the energy is $E_s$, then
the subsequent deterministic trajectory in energy (prior to the next
jump) is given by Equation~(\ref{eq:Et_const_beta0}). Using this
result and Equation~(\ref{eq:lambda_W08}), 
Equation~(\ref{eq:Ftau}) evaluates to
\begin{eqnarray}\label{eq:Ftau_flarelike}
F(\tau ) =\ln (1-u) &+& \frac{\alpha_0/\beta_0}{\gamma-1}
  \left\{ \frac{E_c^{-\gamma+1}}{\delta+1}
  \left[\left(E_s+\beta_0\tau\right)^{\delta+1}
  -\epsilon^{\delta+1}\right]\right.
  \nonumber \\
  &-&\left. \frac{1}{\delta-\gamma+2}
  \left[\left(E_s+\beta_0\tau\right)^{\delta-\gamma+2}
  -\epsilon^{\delta-\gamma+2}
  \right]\right\},
\end{eqnarray}
where $\epsilon=E_s$ if $E_s\geq E_c$, and $\epsilon=E_c$ if $E_s<E_c$.
In this case numerical root finding is required, and 
for the application of Newton-Raphson
it is helpful to note from Equation~(\ref{eq:Ftauprime}) that
\begin{equation}\label{eq:Fprime_flarelike}
F^{\prime}(\tau )=\frac{\alpha_0}{\gamma-1}
\left(E_s+\beta_0\tau\right)^{\delta}
\left[E_c^{-\gamma+1}-\left(E_s+\beta_0\tau \right)^{-\gamma+1}\right].
\end{equation}
The distribution of jump energies defined by Equation~(\ref{eq:h}) is
\begin{equation}
h(\Delta E,E) =
\frac{(\gamma-1)(\Delta E )^{-\gamma+1}}{E_c^{-\gamma+1}-E^{-\gamma+1}}
\end{equation}
for $E_c\leq \Delta E\leq E$. A variable with this distribution may be 
generated from a uniform deviate $u$ for a given $E$ using the 
transformation
\begin{equation}
\Delta E=\left[E_c^{-\gamma+1}
-u\left(E_c^{-\gamma+1}-E^{-\gamma+1}\right)\right]^{-1/(\gamma -1)}.
\end{equation}
A suitable starting energy for a simulation is provided by 
Equation~(\ref{eq:ebar_pl_alpha}).

Following \inlinecite{1998ApJ...494..858W} and 
\inlinecite{2008ApJ...679.1621W}, it is useful to non-dimensionalize, 
by introducing 
\begin{equation}\label{eq:non_dim}
\overline{E}=\frac{E}{E_c}, \quad \overline{t}=\frac{\beta_0t}{E_c}, 
\end{equation}
and
\begin{equation}\label{eq:rbar_def}
\overline{r}=\frac{\alpha_0E_c^{\delta-\gamma+2}}{\beta_0}.
\end{equation}
Equations~(\ref{eq:Ftau_flarelike}) 
and~(\ref{eq:Fprime_flarelike}) become
\begin{eqnarray}\label{eq:FtauWG98_ndim}
F(\overline{\tau }) =\ln (1-u) &+& \frac{\overline{r}}{\gamma-1}
  \left\{ \frac{1}{\delta+1}
  \left[\left(\overline{E}_s+\overline{\tau}\right)^{\delta+1}
  -\overline{\epsilon}^{\delta+1}\right]\right.
  \nonumber \\
  &-&\left. \frac{1}{\delta-\gamma+2}
  \left[\left(\overline{E}_s+\overline{\tau}\right)^{\delta-\gamma+2}
  -\overline{\epsilon}^{\delta-\gamma+2}
  \right]\right\}
\end{eqnarray}
where $\overline{\epsilon}=\overline{E}_s$ if $\overline{E}_s\geq 1$, 
and $\overline{\epsilon}=1$ if $\overline{E}_s< 1$, and 
\begin{equation}
F^{\prime}(\overline{\tau })=\frac{\overline{r}}{\gamma-1}
\left(\overline{E}_s+\overline{\tau}\right)^{\delta}
\left[1-\left(\overline{E}_s+\overline{\tau}\right)^{-\gamma+1}\right],
\end{equation}
respectively.
The transformation used to generate jump energies is
\begin{equation}
\Delta \overline{E} =\left[1
-u\left(1-\overline{E}^{-\gamma+1}\right)\right]^{-1/(\gamma -1)},
\end{equation}
and the starting energy is
\begin{equation}\label{eq:Emean_nd}
\langle \overline{E}\rangle \approx
\left(\frac{2-\gamma}{\overline{r}}
\right)^{1/(\delta+2-\gamma )}.
\end{equation}

The parameters $\delta$ and $\overline{r}$ define the specific
model being solved. Equation~(\ref{eq:Emean_nd}) implies that 
$\overline{r}<\frac{1}{2}$ is required for 
$\langle \overline{E}\rangle>1$. Smaller values of $\overline{r}$ 
lead to larger values of $\langle \overline{E}\rangle$ and hence 
more decades of power-law behavior in the frequency-energy
distribution, as explained in Section~\ref{sec:Master-Flare-like}.
Many decades of power-law behavior are observed for flares on
the Sun, implying a small value of $\overline{r}$. 

\begin{figure}[here]
\centerline{\includegraphics[width=0.75\textwidth]{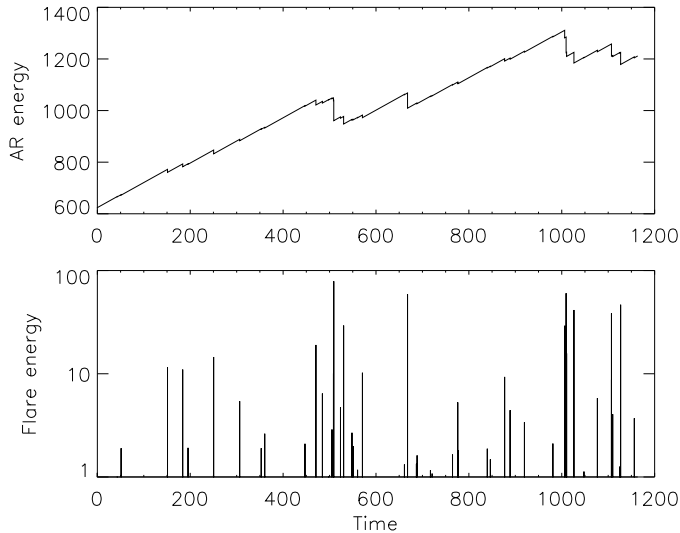}}
\caption{Monte-Carlo solution of the flare-like case from
Wheatland and Glukhov (1998), with $\delta=0$ and $\overline{r}=0.02$.
The upper panel shows the active-region energy {\it versus} time, and 
the lower panel shows the flare energy {\it versus} time, for a period of
time including 50 jump transitions.}
\label{fig:fig3}
\end{figure}

As an example of applying 
Equations~(\ref{eq:FtauWG98_ndim})\,--\,(\ref{eq:Emean_nd}), we 
consider the case $\delta=0$, from \inlinecite{1998ApJ...494..858W}. 
Figure~\ref{fig:fig3} shows the results of a Monte-Carlo solution with
$\delta=0$ and $\overline{r}=0.02$. The upper panel shows the time 
history of the energy of the model active region over the first 50 
jumps, and the lower panel shows the corresponding flare energies 
{\it versus} 
time (with a logarithmic scale). The mean energy of the system is
$\langle \overline{E}\rangle = 625$, which is the starting energy
for the simulation. Figure~\ref{fig:fig3} illustrates the character
of the flare-like models. The active-region energy grows linearly with
time between flares, flare sizes are power-law distributed, and
the total flaring rate is approximately constant, since 
$\overline{E}\gg 1$. During this period of time only relatively small 
flares occurred, with the largest event having energy close to 100 in 
non-dimensional units.

\begin{figure}[here]
\centerline{\includegraphics[width=1\textwidth]{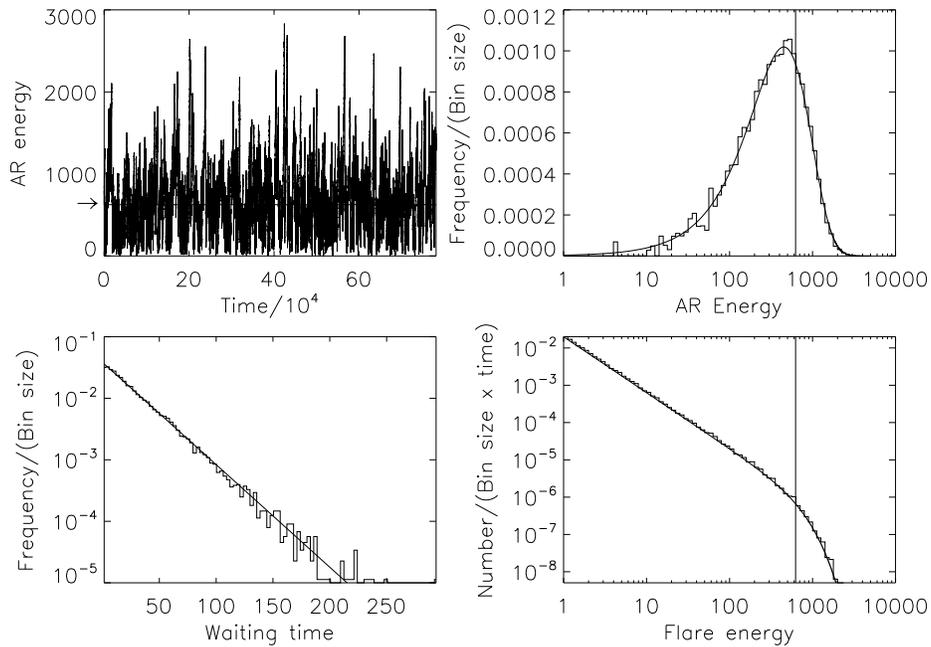}}
\caption{Monte-Carlo solution of the flare-like case from
Wheatland and Glukhov (1998), involving $3\times 10^4$ jump 
transitions, and 
the parameter value $\overline{r}=0.02$. The figure shows the 
active-region energy {\it versus} time (upper left), and histograms of:
the active-region energy (upper right); the waiting-time distribution 
(lower left); and the flare frequency-energy distribution (lower
right). The corresponding results obtained by solving the master 
equation in the steady state are shown 
by solid curves. The active-region mean energy is indicated by an arrow
near the left-hand axis of the upper-left panel, and is shown by a 
solid vertical line in the two panels on the right.}
\label{fig:fig4}
\end{figure}

Figure~\ref{fig:fig4} illustrates the Monte-Carlo 
solution with the same parameters ($\delta=0$ and $\overline{r}=0.02$) 
for $3\times 10^4$ waiting times and jump transitions, and compares 
the results with direct solution of the master equation.
The upper-left panel shows the time history of the energy of the 
system, with the mean energy $\langle \overline{E}\rangle = 625$ 
shown by a horizontal line, which is also indicated by an arrow near
the left-hand axis of the panel. The upper-right panel shows the 
histogram of the energy of the system, obtained by sampling the 
simulated time history of energy at $3\times 10^4$ random times. 
The mean energy is shown by a solid vertical line. The solid curve is 
the distribution obtained by solving the master equation in the steady 
state, using the method in \inlinecite{2008ApJ...679.1621W}. The 
lower-left panel shows the histogram of waiting times, in a log-linear
representation, together with the 
waiting-time distribution obtained from the solution to the master 
equation (solid curve). The lower-right panel shows the flare 
frequency-energy histogram, together with the distribution obtained 
from the solution to the master equation (solid curve), and the mean 
energy (solid vertical line). The Monte-Carlo solutions agree with the 
direct solution of the master equation. These results confirm the 
finding in \inlinecite{2008ApJ...679.1621W} that the waiting-time 
distribution is essentially exponential in this case. 

The comparison in Figure~\ref{fig:fig4} illustrates the 
computational advantage of the Monte-Carlo method over direct
solution of the master equation. The results shown by
the solid curves in Figure~\ref{fig:fig4} require the solution of 
a linear system in 5000 unknowns, to obtain sufficient energy 
resolution to ensure accuracy. By comparison, the Monte-Carlo solution
requires the simulation of $3\times 10^4$ waiting times and jump
transitions, but each of these calculations is simple. As a result
the Monte-Carlo method is substantially faster than the linear
solution. 

\subsection{WAITING-TIME DISTRIBUTIONS FOR FLARE-LIKE CASES} 
\label{sec:SDE-Flare-like-WTD}

In \inlinecite{2008ApJ...679.1621W} it was shown using numerical
solutions of the master equation that the flare-like models with 
$\delta=0$ and $\delta=1$ exhibit approximately Poisson (exponential) 
waiting-time distributions, for certain choices
of the non-dimensional ratio 
$\overline{r}=\alpha_0E_c^{\delta-\gamma+2}/\beta_0$ of transition 
rates to energy-supply rate [in \inlinecite{2008ApJ...679.1621W} this 
ratio was labelled $\overline{\alpha}_0$, or just $\alpha_0$, when the
bar was dropped]. These results are briefly discussed in 
Section~\ref{sec:Master-Flare-like}. However, the results also showed
some departure from the simple Poisson distribution. In this section
we further investigate the behavior of the models and in 
particular the waiting-time distributions, using Monte-Carlo 
solutions.

\begin{figure}
\centerline{\includegraphics[width=0.33\textwidth]{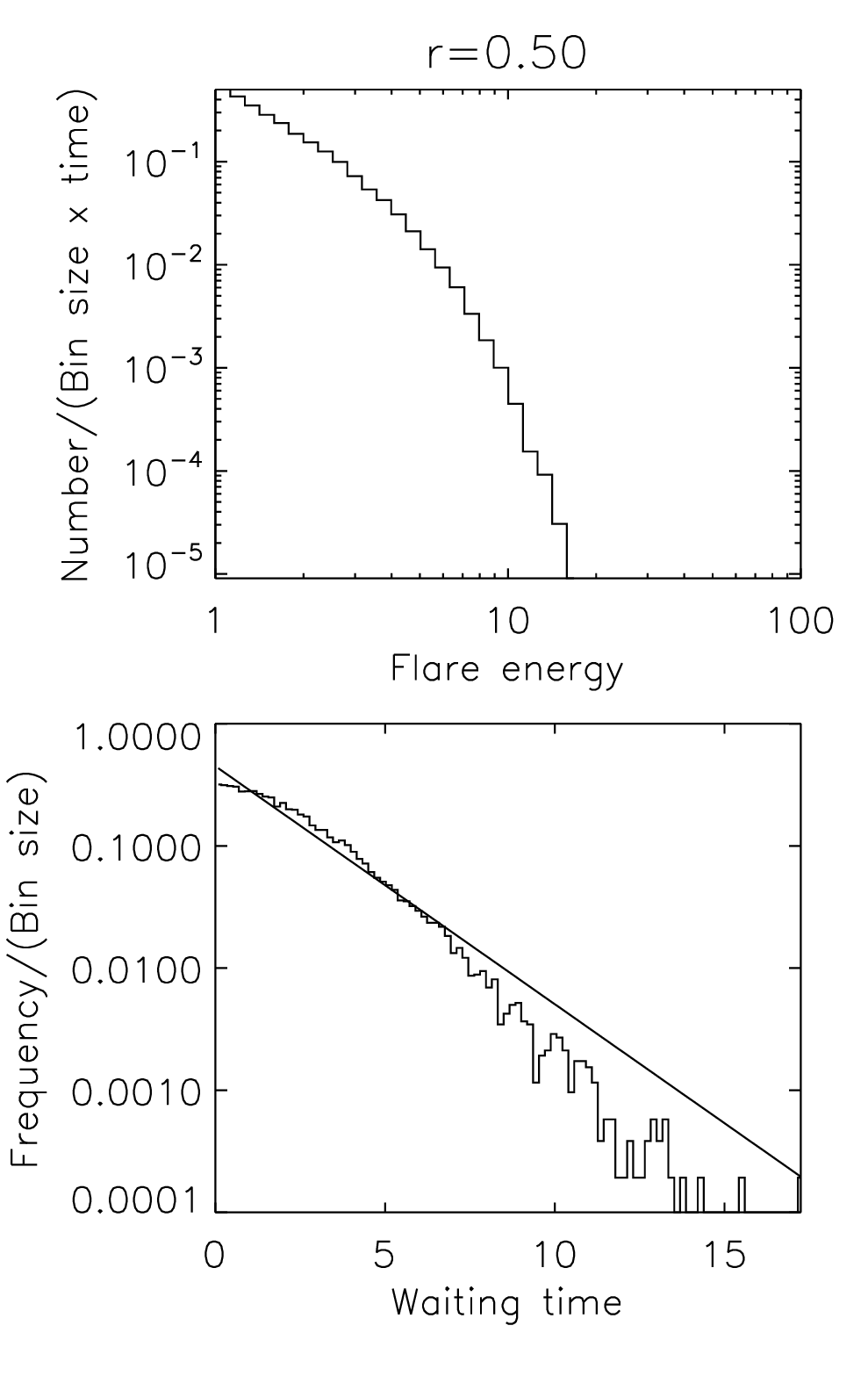}
            \includegraphics[width=0.33\textwidth]{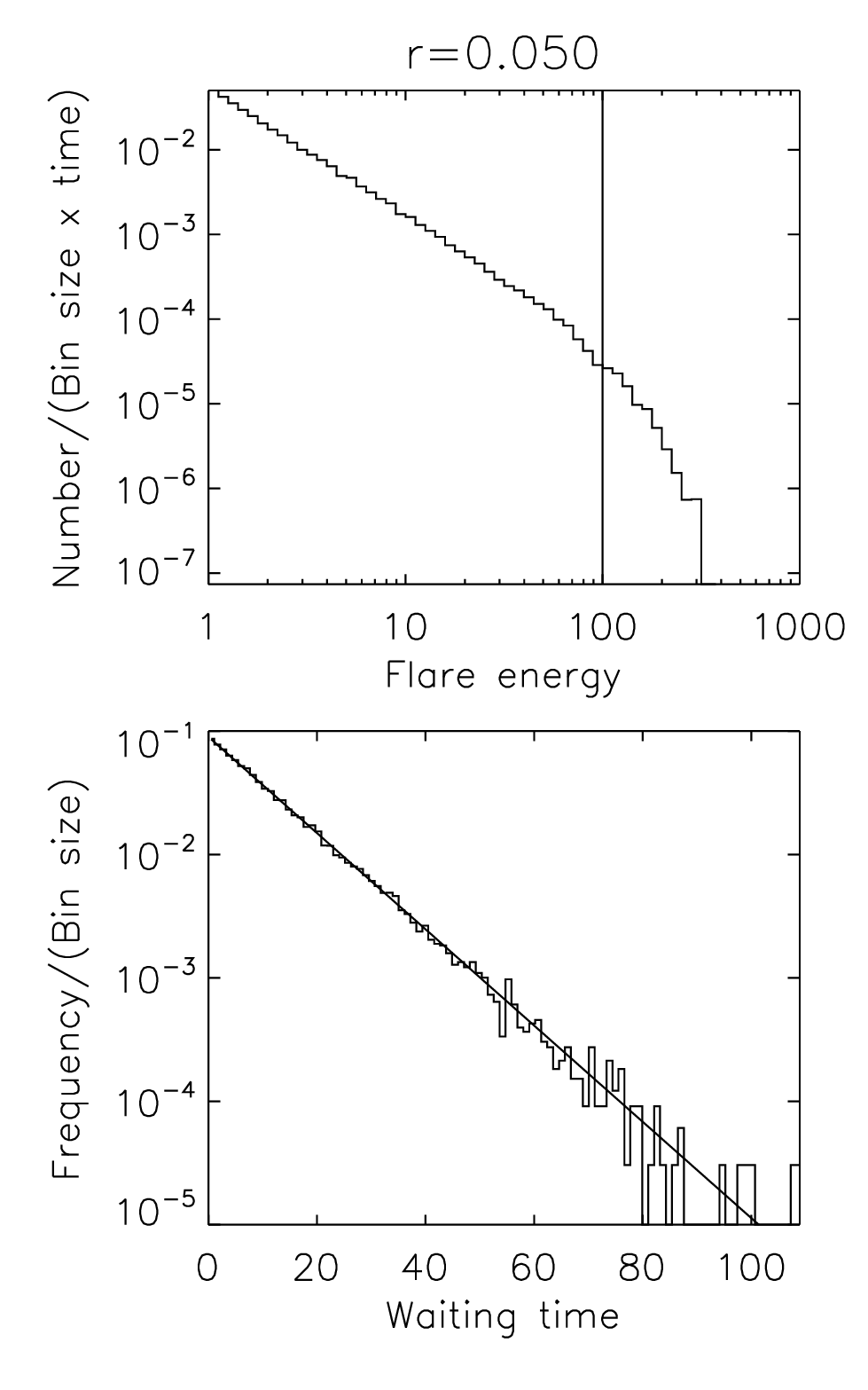}
            \includegraphics[width=0.33\textwidth]{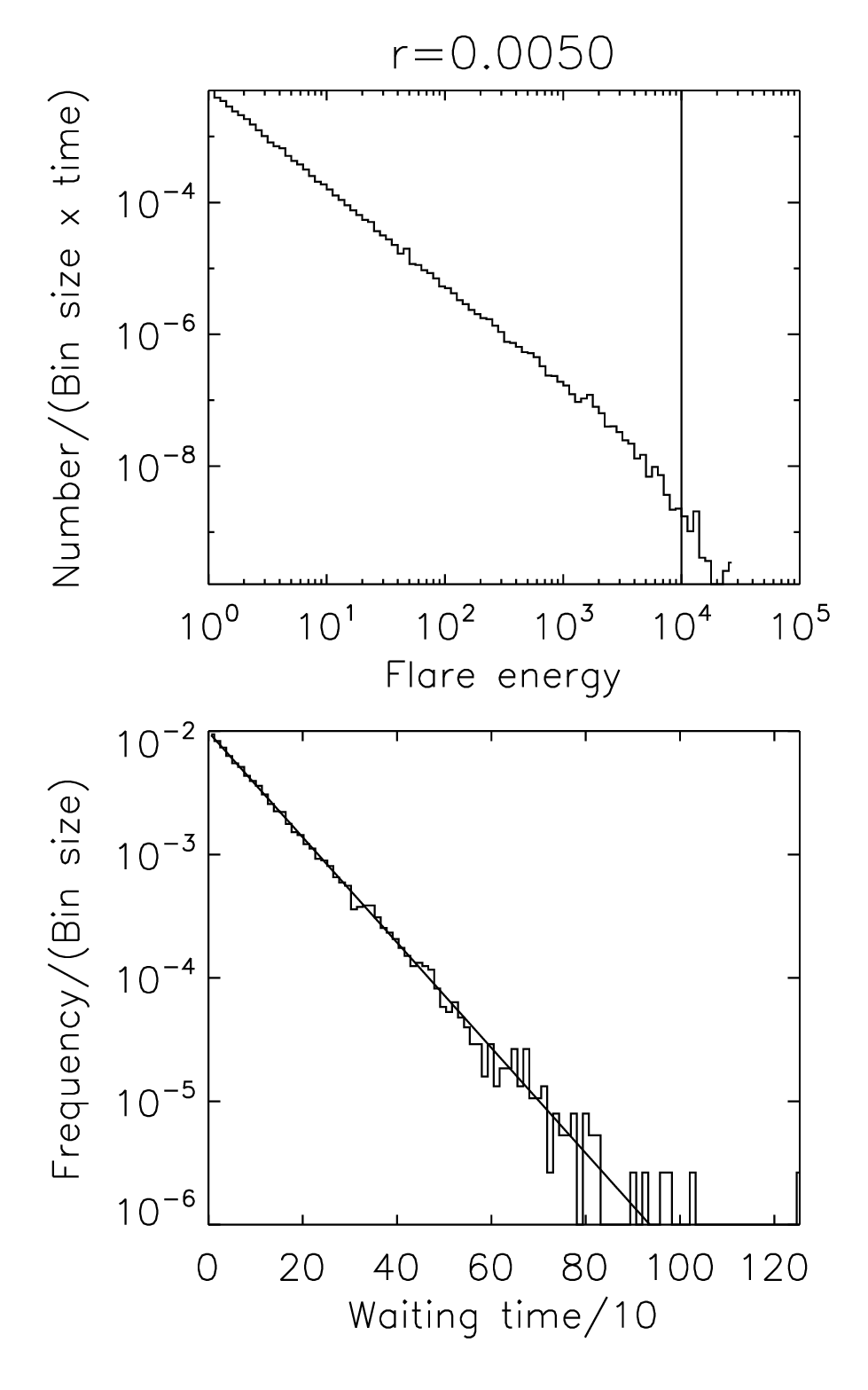}}
\caption{The flare frequency-energy distributions (upper row) and 
flare waiting-time distributions (lower row) for the flare-like case 
with $\delta=0$. The left-hand pair of distributions is for
$\langle \overline{E}\rangle \approx 1$
($\overline{r}=0.5$), the center pair is for 
$\langle \overline{E}\rangle \approx 10^2$ 
($\overline{r}=5\times 10^{-2}$), and
the right-hand pair is for 
$\langle \overline{E}\rangle \approx 10^4$
($\overline{r}=5\times 10^{-3}$).}
\label{fig:fig5}
\end{figure}

Figure~\ref{fig:fig5} illustrates three solutions for the case 
$\delta=0$. Each solution involves $3\times 10^4$ waiting times and 
jump transitions. The upper row in Figure~5 shows the flare 
frequency-energy distributions in a log-log representation,
and the lower row shows the corresponding waiting-time distributions,
in a log-linear representation.
The left-hand pair of distributions is for $\overline{r}=0.5$, the
center pair is for $\overline{r}=0.05$, and the right-hand pair is
for $\overline{r}=0.005$. The solid vertical lines in the 
frequency-energy distributions show the approximation to the mean 
energy $\langle \overline{E}\rangle$ 
given by Equation~(\ref{eq:Emean_nd}). The three choices of 
$\overline{r}$ shown correspond to values 
$\langle \overline{E}\rangle =1$, $\langle \overline{E}\rangle =10^2$, 
and $\langle \overline{E}\rangle =10^4$. The solid lines on the 
waiting-time distributions show the exponential form 
$\overline{\lambda}_m e^{-\overline{\lambda}_m\overline{\tau}}$, where 
$\overline{\lambda}_m$ is the overall
mean rate of events, {\it i.e.} the number of events divided by the 
total time (non-dimensionalised). The upper row of Figure~5 confirms 
that the frequency-energy 
distribution is a power law with index $\gamma$ below a roll-over set 
by the largest energy the system is likely to attain, which may be
roughly approximated by $\langle \overline{E}\rangle$. For 
smaller values of $\overline{r}$, 
the system attains larger energies, as flaring is less frequent. 
The lower row of Figure~\ref{fig:fig5} shows that the waiting-time 
distribution becomes exponential as $\langle \overline{E}\rangle$ 
increases (or as $\overline{r}$ decreases). 
This may be understood using the argument given
in Section~\ref{sec:Master-Flare-like}: for $\overline{r}\ll 1$, 
the energy $\overline{E}$ of the system satisfies $\overline{E}\gg 1$, 
in which case the approximation of Equation~(\ref{eq:lambda_W08_largeE}) 
applies, and the total rate of flaring $\overline{\lambda} (\overline{E})$ 
is then independent of $\overline{E}$.

\begin{figure}
\centerline{\includegraphics[width=0.33\textwidth]{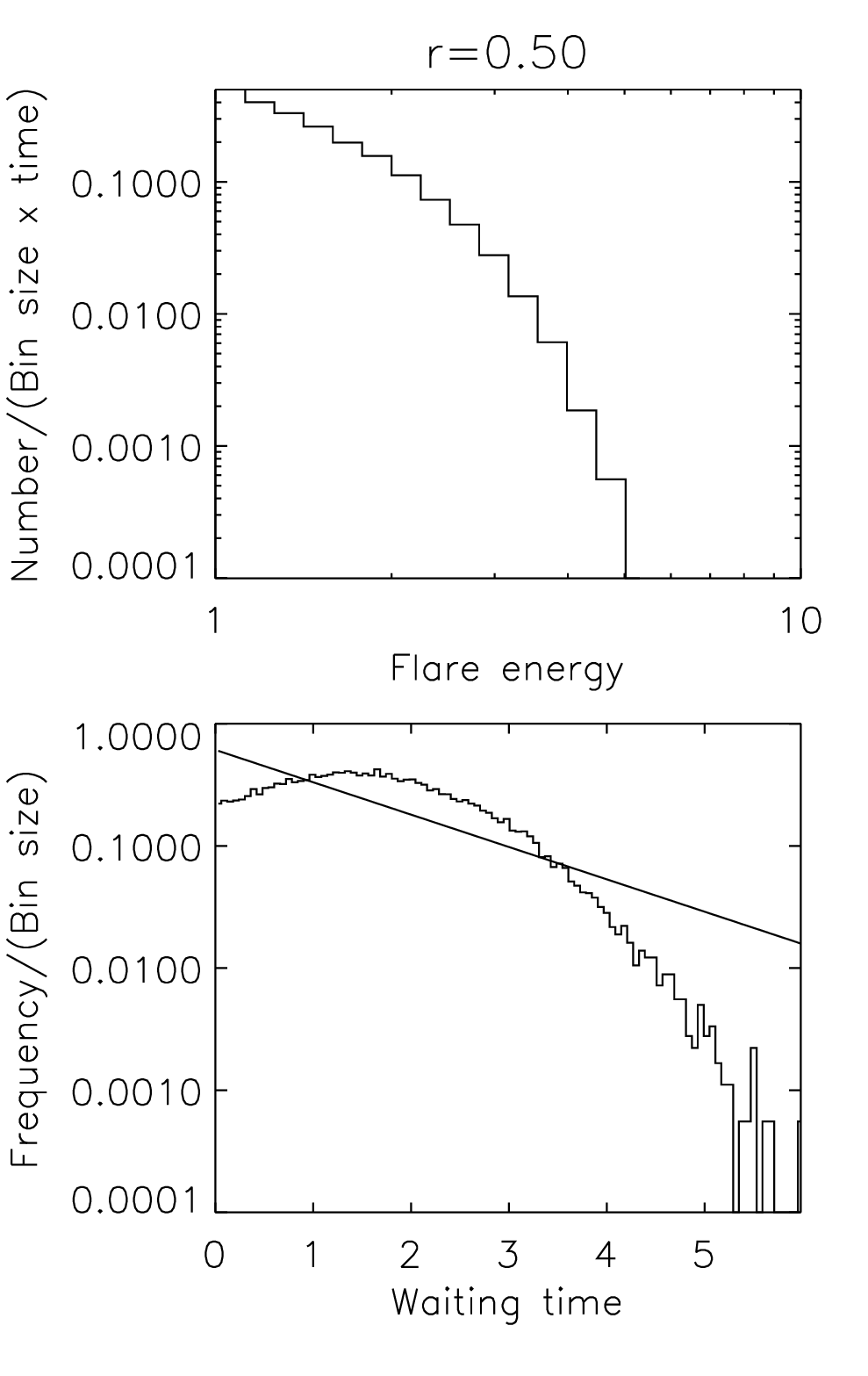}
            \includegraphics[width=0.33\textwidth]{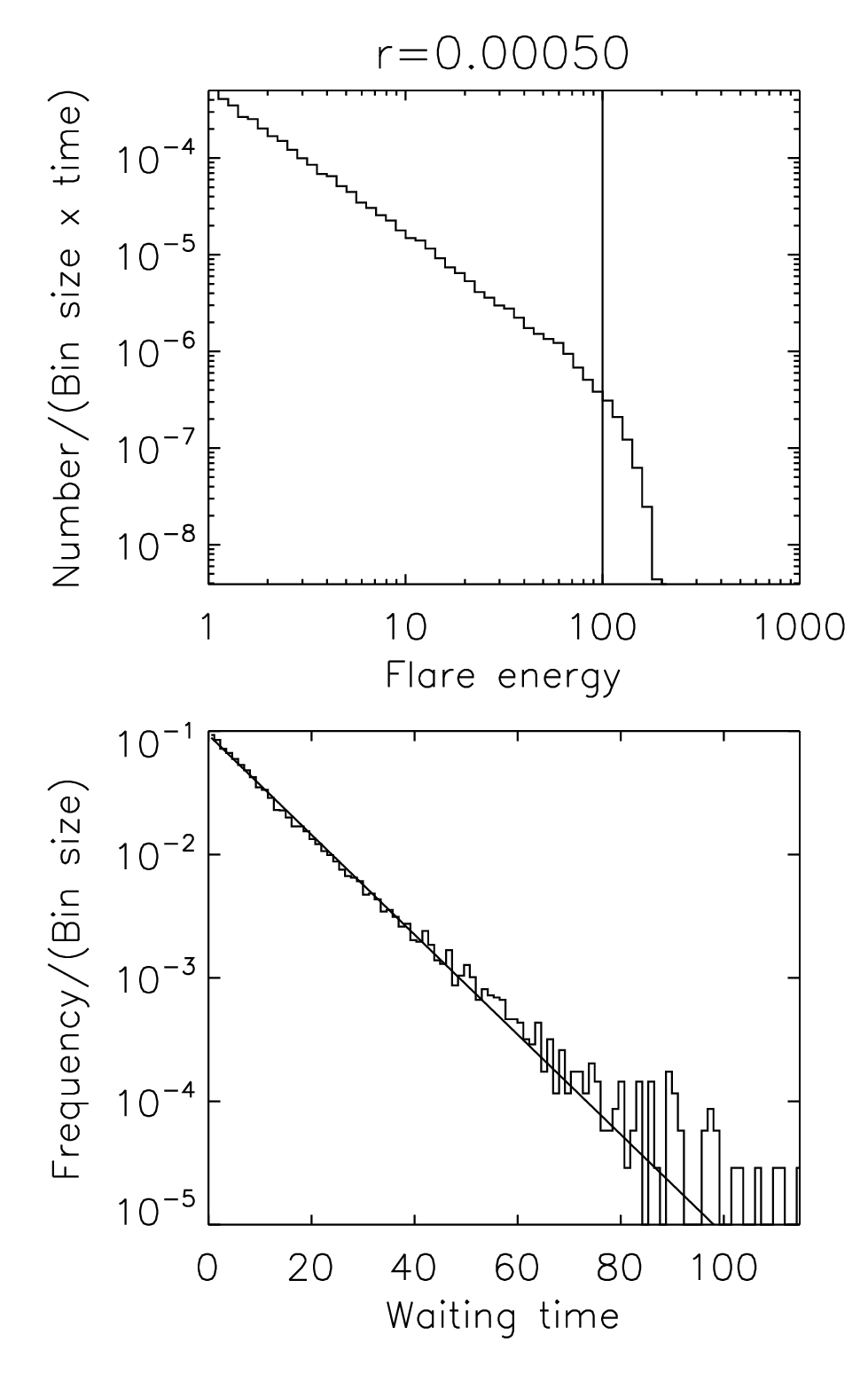}
            \includegraphics[width=0.33\textwidth]{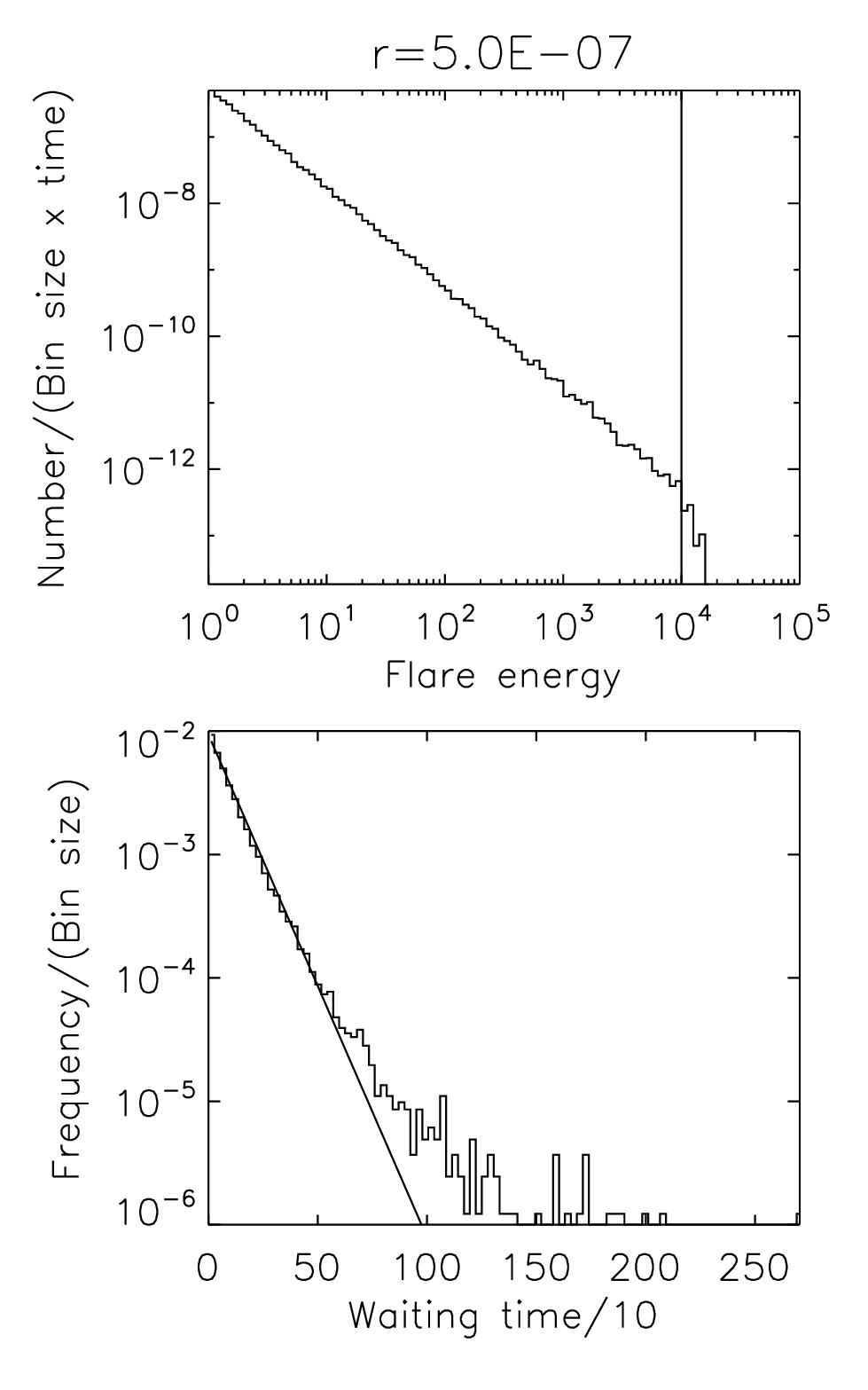}}
\caption{The flare frequency-energy distributions (upper row) and flare
waiting-time distributions (lower row) for the flare-like case with 
$\delta=1$. The left-hand pair of distributions is for
$\langle \overline{E}\rangle = 1$
($\overline{r}=0.5$), the center pair is for 
$\langle \overline{E}\rangle = 10^2$ 
($\overline{r}=5\times 10^{-4}$), and
the right-hand pair is for 
$\langle \overline{E}\rangle = 10^4$
($\overline{r}=5\times 10^{-7}$).}
\label{fig:fig6}
\end{figure}

Figure~\ref{fig:fig6} illustrates three solutions for the case 
$\delta = 1$, again with $3\times 10^4$ waiting times and 
jump transitions. The format of the figure is the same as for
Figure~\ref{fig:fig5}, and the values of $\overline{r}$ for the three
cases are chosen so that the values of $\langle \overline{E}\rangle$ 
[using the approximation of Equation~(\ref{eq:Emean_nd})] are the
same, {\it i.e.} $\langle \overline{E}\rangle =1$ (left), 
$\langle \overline{E}\rangle =10^2$ (center), 
and $\langle \overline{E}\rangle =10^4$ (right). The corresponding 
values of $\overline{r}$ are given above the three rows in the figure.
The upper row of 
Figure~\ref{fig:fig6} shows the expected power-law behavior below
an upper roll-over given approximately by 
$\langle \overline{E}\rangle$. The lower row of Figure~\ref{fig:fig6} 
shows that the waiting-time distribution is approximately exponential 
for $\langle \overline{E}\rangle = 10^2$, but for 
$\langle \overline{E}\rangle = 10^4$ there is an excess of large 
waiting times by comparison with the
exponential form. Referring again to the argument in 
Section~\ref{sec:Master-Flare-like}, for $\overline{r}\ll 1$ we 
expect $\overline{E}\gg 1$, in which case
the total flaring rate varies approximately as
$\overline{\lambda} (\overline{E}) \propto \overline{E}$. The 
linear variation in 
mean flaring rate with energy leads to some departure from the simple 
exponential form, with the departure becoming more significant as 
$\overline{r}$ decreases.

\begin{figure}
\centerline{\includegraphics[width=0.33\textwidth]{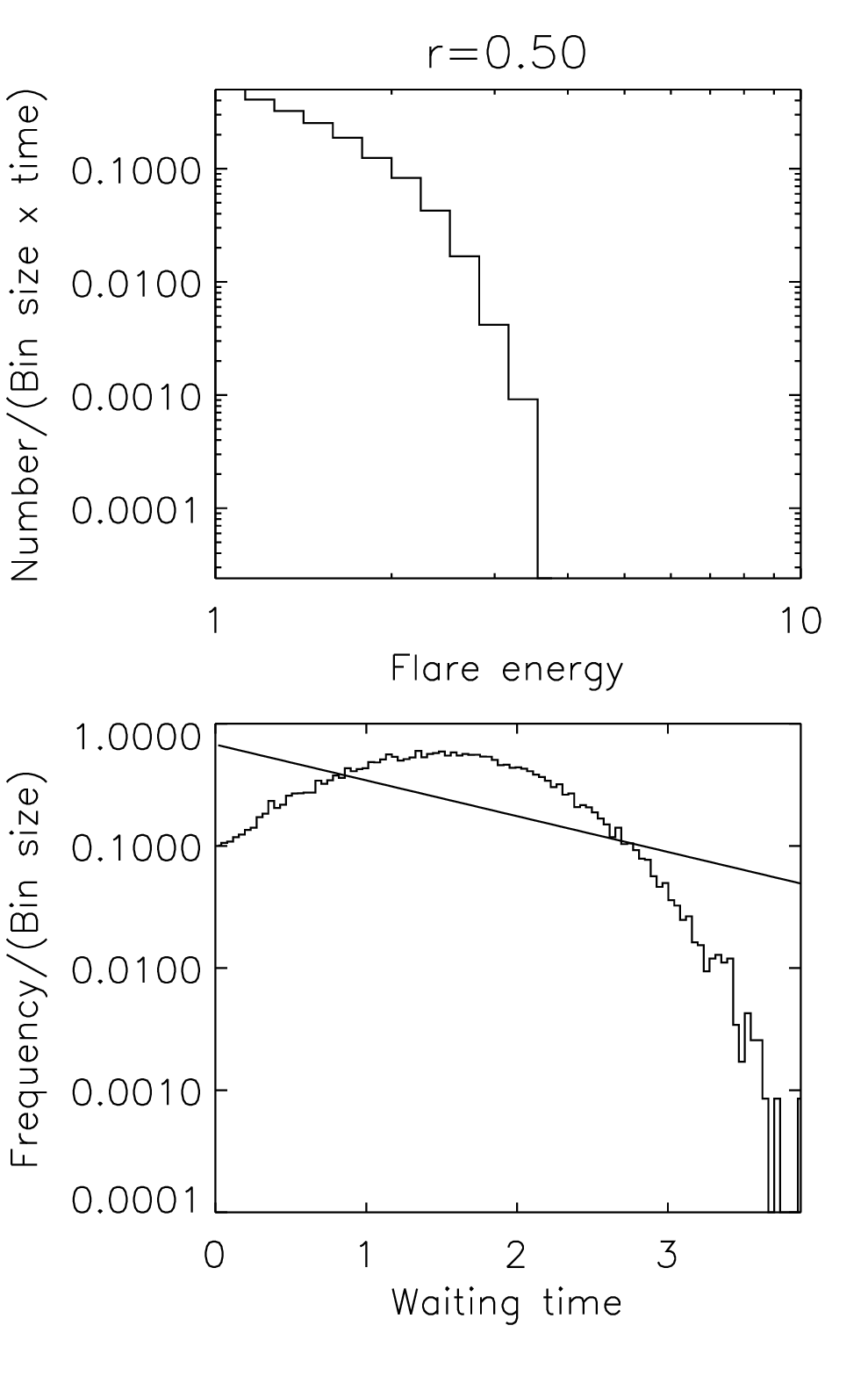}
            \includegraphics[width=0.33\textwidth]{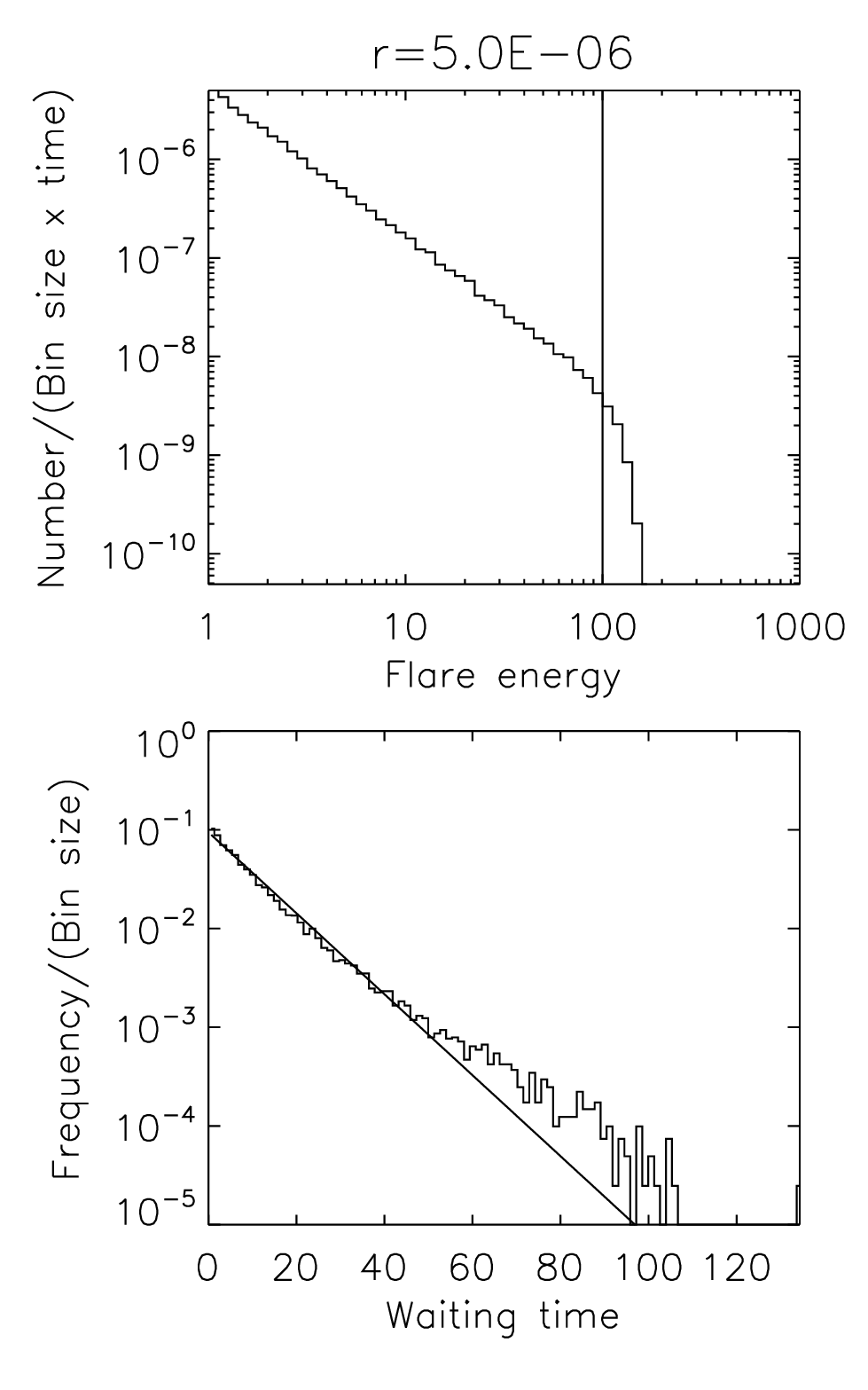}
            \includegraphics[width=0.33\textwidth]{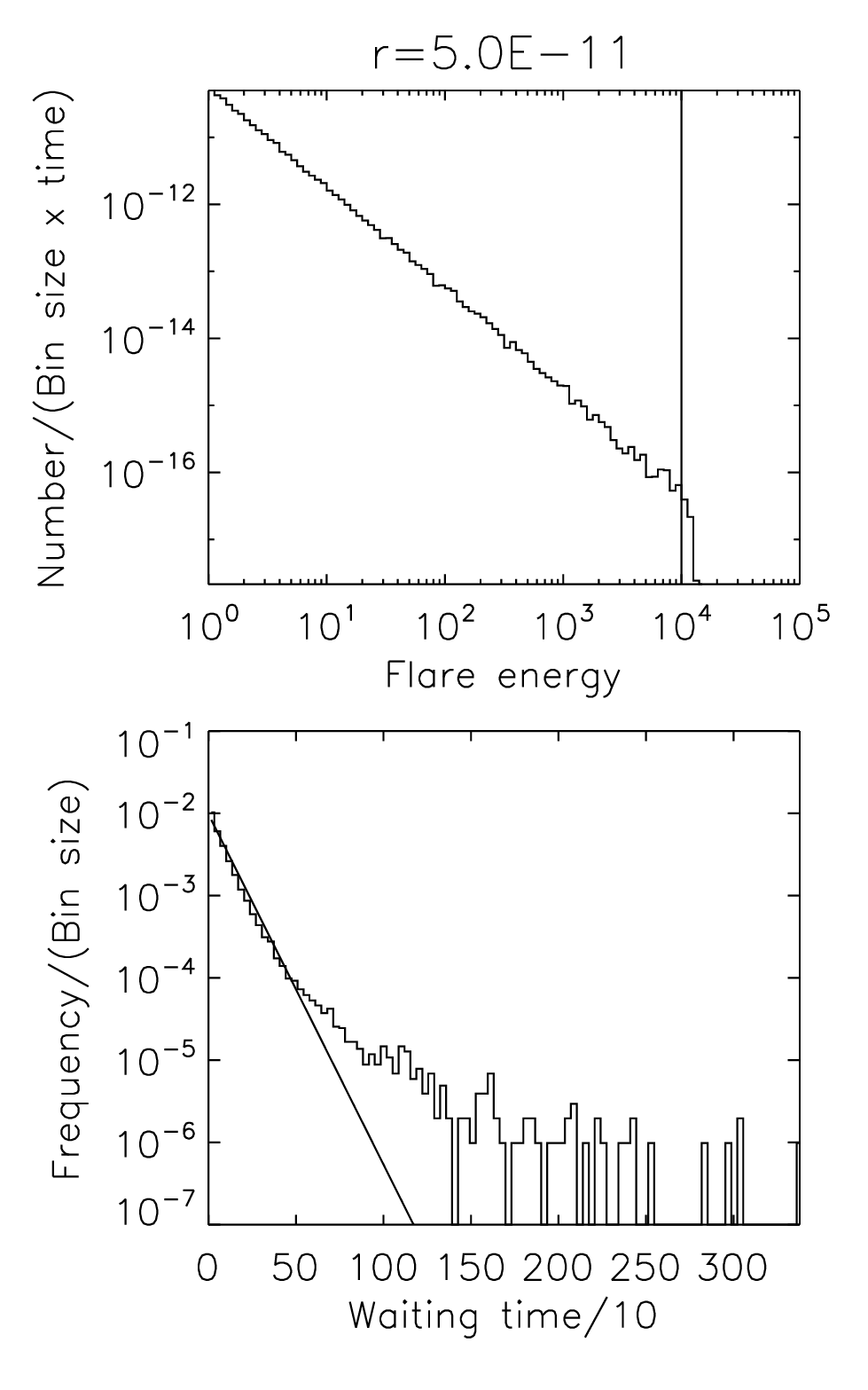}}
\caption{The flare frequency-energy distributions (upper row) and 
flare waiting-time distributions (lower row) for the flare-like case 
with $\delta=2$. The left-hand pair of distributions is for
$\langle \overline{E}\rangle = 1$
($\overline{r}=0.5$), the center pair is for 
$\langle \overline{E}\rangle = 10^2$ 
($\overline{r}=5\times 10^{-6}$), and
the right-hand pair is for 
$\langle \overline{E}\rangle = 10^4$
($\overline{r}=5\times 10^{-11}$).}
\label{fig:fig7}
\end{figure}

Figure~\ref{fig:fig7} illustrates three solutions for the case 
$\delta = 2$, again with $3\times 10^4$ waiting times and 
jump transitions. The format of the figure is the same as for
Figures~\ref{fig:fig5} and~\ref{fig:fig6}, with the values of
$\overline{r}$ again corresponding to approximate mean energies
$\langle \overline{E}\rangle =1$ (left), 
$\langle \overline{E}\rangle =10^2$ (center), 
and $\langle \overline{E}\rangle =10^4$ (right).
The results for the 
frequency-energy distribution (upper row) are as expected, with
power-law behavior below an upper roll-over given approximately 
by $\langle \overline{E}\rangle$. The results for the waiting-time
distribution are similar to the case $\delta=1$, but are more 
pronounced.
For $\langle \overline{E}\rangle = 10^2$ there is very approximate
exponential behavior, with some excess of large waiting times. For 
$\langle \overline{E}\rangle = 10^4$ there is substantial
departure from the exponential model, with a pronounced excess 
of large waiting times. 
Using the argument in Section~\ref{sec:Master-Flare-like}, we expect
approximate $\propto \overline{E}^2$ dependence of the total flaring 
rate for $\overline{E}\gg 1$. This variation in the mean flaring rate
with energy leads to the departure from the exponential form.

The results shown in Figures~\ref{fig:fig5}\,--\,\ref{fig:fig7} 
suggest that
the $\delta = 0 $ model may be the preferred flare-like solution. 
The flare frequency-energy distribution is observed to be a power law
over many decades in energy, which implies $\overline{r}\ll 1$. The
models with $\delta \neq 0$ exhibit significant departure from 
Poisson waiting-time statistics for $\overline{r}\ll 1$. However, it 
is not clear that any such departure is observed for flares on the Sun:
some active regions that produce very large flares appear to exhibit
simple Poisson waiting-time statistics
\cite{2001JGR...10629951M,2001SoPh..203...87W}. 

\begin{figure}[here]
\centerline{\includegraphics[width=0.75\textwidth]{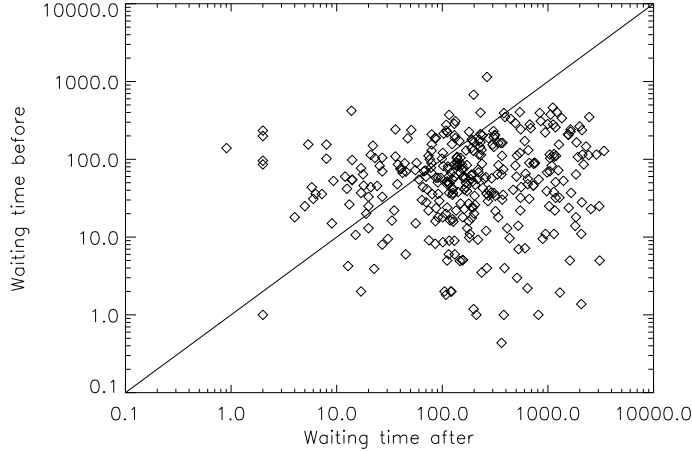}}
\caption{Plot of the waiting time before an event, {\it versus} 
the waiting time after an event, for all events larger than 
$0.2\langle\overline{E}\rangle$, for the simulation with $\delta=2$
and $\langle\overline{E}\rangle = 10^4$ shown in 
Figure~\ref{fig:fig7}.}
\label{fig:fig8}
\end{figure}

The variation in total flaring rate 
with energy of the system for the cases with $\delta\neq 0$ also 
implies observable consequences. For example, 
if we pick out ``large'' events from a simulation ensemble 
with $\delta\neq 0$ and plot the waiting time
before the event {\it versus} the waiting time after the event, then we 
expect that the waiting times after the event will tend to be larger
than the waiting times before, since large events deplete the system
energy, and hence reduce the total rate. Figure~\ref{fig:fig8} illustrates
this effect. The figure is constructed using the simulation shown in 
Figure~\ref{fig:fig7} with $\delta=2$ and 
$\langle \overline{E}\rangle = 10^4$, and the threshold for a 
large event is chosen to be $0.2\langle \overline{E}\rangle$. The solid 
line is the line of equality of the two waiting times. As expected, the
waiting times after the large events tend to be larger than the waiting 
times before the event (more points lie below the line of equality than
above). No such effect is observed when a similar plot 
is constructed for the model with $\delta=0$ and 
$\langle \overline{E}\rangle = 10^4$.
This effect is not observed for flares
on the Sun \cite{1998ApJ...509..448W,2001SoPh..203...87W}.

\section{Conclusions} 
\label{sec:Conclusions}

This paper presents a Monte-Carlo method for solving the stochastic
model for active region energy presented in 
\inlinecite{1998ApJ...494..858W} and \inlinecite{2008ApJ...679.1621W},
which in particular is suited to solving the flare-like cases in 
those papers. The method
numerically solves the stochastic differential equation describing the
system, rather than the equivalent master equation, and provides a 
computationally-efficient approach to the problem. The 
method is demonstrated on a simple Gaussian test, and is compared with
a direct solution of the steady-state master equation for a 
flare-like case from \inlinecite{1998ApJ...494..858W}.

The method is used to further investigate the class of flare-like 
models from \inlinecite{2008ApJ...679.1621W}, which feature constant 
energy-supply rates $\beta_0$ and flare transition rates
of the form 
\begin{equation}\label{eq:pl_alpha_whe08_rep}
\alpha (E,E^{\prime})=\alpha_0 E^{\delta} 
(E-E^{\prime})^{-\gamma}\theta (E-E^{\prime}-E_c),
\end{equation}
where $\alpha_0$ is a constant, and $E_c$ is a low-energy cutoff.
The index $\gamma=1.5$ is the observed flare frequency-energy power
law index, and $\delta$ is a positive constant (we consider the cases
$\delta =0$, $\delta =1$, and $\delta =2$).
The emphasis in the investigation is on the waiting-time distributions 
for these models, and their adherance to/departure from a Poisson
(exponential) form, as a function of $\delta$ and of the dimensionless
ratio $\overline{r}=\alpha_0E_c^{\delta-\gamma+2}/\beta_0$.
The models require small values of $\overline{r}$ to produce flares
with a frequency-energy distribution exhibiting a power law over many
decades [a lower bound to the departure from power-law behavior is set 
by the estimate of the mean energy $\langle \overline{E}\rangle \approx
\left[(2-\gamma)/\overline{r}\right]^{1/(\delta+2-\gamma )}$]. For
the $\delta =0$ model it is found that the waiting-time distribution
becomes a close approximation to a simple exponential for small 
$\overline{r}$. This may be explained in terms of the total rate of
flaring becoming constant for large $\overline{E}$, which applies
when $\overline{r}\ll 1$. For the $\delta =1$ and $\delta = 2$ models 
the waiting-time distribution is approximately exponential for 
intermediate values of $\overline{r}$, but exhibits an excess of
large waiting times for $\overline{r}\ll 1$. This result may make the
$\delta = 0$ case the preferred flare-like solution, since it is unclear
that any such departure is observed for flares on the Sun. Also, the
dependence of the total flaring rate on the active-region energy for the
$\delta=1$ and $\delta=2$ models implies variations in flare rate with 
flare occurrence which do not appear to be observed on the Sun.

It is interesting to reconsider the correspondence between the 
stochastic model and the avalanche model for
flares~\cite{1991ApJ...380L..89L,2001SoPh..203..321C}, in light of
the new results. The 
investigation in this paper excludes the possibility of a stochastic
model with transition rates of the form of 
Eq.~(\ref{eq:pl_alpha_whe08_rep}) with $\delta\neq 0$, which 
simultaneously exhibits a wide 
range of flare energies and has uncorrelated waiting times. This is 
due to large flares changing the system energy and hence total 
flaring rate significantly. The choice for the transition
rates was motivated in part by the avalanche model (as outlined in
Section~\ref{sec:Master-Flare-like}). Avalanche models exhibit 
power-law frequency-energy distributions over many decades, and they 
have uncorrelated waiting times~\cite{1998ApJ...509..448W}.
However, in avalanche models the largest flares deplete only a
small fraction of of the avalanche grid ``energy'' (see {\it e.g.}
Fig.~3 in \opencite{2001SoPh..203..321C}). The discrepancy between 
the two pictures might be due to different definitions of energy 
(the avalanche model ``energy'' may correspond to free energy plus 
background magnetic energy unavailable for flares). Another 
possibility is that the $\delta=0$ model provides the closest 
match to the avalanche picture. A more careful analysis of the 
correspondence between the two models is needed to resolve this 
point. 

The Monte-Carlo approach has several advantages over solution of the
master equation. First, as discussed in Section~\ref{sec:SDE-Flare-like},
the Monte-Carlo solution is numerically simple to implement and
computationally efficient. Second, as highlighted by Figure~\ref{fig:fig8},
since the Monte-Carlo approach produces an ensemble of flare events,
it permits detailed investigation of event statistics. A third advantage 
of the Monte-Carlo method is that it permits solution of the model in 
time-dependent situations. In principle solutions may be constructed 
for arbitrary time variation of the rates $\beta (E,t)$ and 
$\alpha (E,E^{\prime},t)$. The methods of solution of the master
equation presented in \inlinecite{1998ApJ...494..858W} and 
\inlinecite{2008ApJ...679.1621W} apply only in the steady
state, and in particular the method of obtaining the waiting-time 
distribution requires a steady state. As discussed in 
Section~\ref{sec:Introduction}, time variation in observed flaring
rates plays a role in determining the observed waiting-time 
distribution, so it is of interest to consider time-dependent models.
These models will be addressed in future work. 

It is possible to construct more general stochastic
models for active-region energy. Active regions may lose energy by
mechanisms other than flaring, {\it e.g.} due to slow ohmic dissipation
of electric currents. If the energy loss is considered to be 
deterministic, and smaller in magnitude than the energy input rate, 
then $\beta (E,t)$ in the present model may be interpreted as 
a {\it net} energy-supply rate, and the model may be considered to 
already accommodate energy losses of this kind. If the energy loss is 
assumed to occur via small random decrements, then a suitable model 
may be to include a Fokker-Planck (diffusion) term in the master 
equation, describing continual small random increases and decreases 
in energy. (The increases may correspond to fluctuations in the 
energy supply rate.) A more general form for the jump-transition 
master equation, sometimes called the Chapman-Kolmogorov 
equation~\cite{Gardiner2004} includes drift (representing deterministic
energy input or loss), diffusion (small random input or loss), and 
jump transitions. An active-region model of this form was briefly 
discussed in \inlinecite{2008ApJ...679.1621W}. In the context 
of the Monte-Carlo approach, the diffusion term corresponds to a Wiener
process, and a different method of solution of the stochastic DE is 
then required. 

Models of the kind presented in this paper are difficult to test in
detail against flare observations because of the difficulties associated
with determining the free energy of active regions, and the rate of 
energy supply to active regions. However, we note that improved solar 
observations (and analysis techniques) may eventually provide such 
information. Independent of this, the
models provide important qualitative checks on our understanding of
energy balance in solar active regions. For example, as noted in 
Section~\ref{sec:Master-Flare-like}, it has proven difficult to identify
choices other than the flare-like ones investigated here that produce
suitable power-law flare frequency-energy distributions. The models are
also of intrinsic interest because of their description of a dynamical 
balance involving scale-free transitions, and it is possible that they
provide suitable descriptions of a variety of other physical systems 
exhibiting power-law behavior.

\begin{acks}
The author thanks Dr.\ Alex Judge, Dr.\ Xue Yang, and Professor Don
Melrose for comments on drafts of the paper. An anonymous referee also 
provided constructive comments.
\end{acks}


\end{article} 
\end{document}